\documentclass[sigconf]{acmart}

\AtBeginDocument{%
  \providecommand\BibTeX{{%
    \normalfont B\kern-0.5em{\scshape i\kern-0.25em b}\kern-0.8em\TeX}}}

\copyrightyear{2024}
\acmYear{2024}
\setcopyright{acmlicensed}\acmConference[KDD '24]{Proceedings of the 30th ACM SIGKDD Conference on Knowledge Discovery and Data Mining}{August 25--29, 2024}{Barcelona, Spain}
\acmBooktitle{Proceedings of the 30th ACM SIGKDD Conference on Knowledge Discovery and Data Mining (KDD '24), August 25--29, 2024, Barcelona, Spain}
\acmDOI{10.1145/3637528.3671506}
\acmISBN{979-8-4007-0490-1/24/08}

\usepackage{multirow}
\usepackage{svg}
\usepackage{graphicx}
\usepackage{float}
\usepackage{subfigure}
\usepackage{algorithm}
\usepackage{algpseudocode}

\begin{document}

\title{Future Impact Decomposition in Request-level Recommendations}

\author{Xiaobei Wang $^{\ast\dagger}$}
\thanks{$^\dagger$ Work was done during an internship at Kuaishou Technology.}
\affiliation{%
  \institution{Peking University}
  \city{Beijing}
  \country{China}}
\email{wangxiaobei@stu.pku.edu.cn}

\author{Shuchang Liu $^\ast$}
\thanks{$^\ast$ The first two authors contributed equally to this work.}
\affiliation{%
  \institution{Kuaishou Technology}
  \city{Beijing}
  \country{China}}
\email{liushuchang@kuaishou.com}

\author{Xueliang Wang}
\affiliation{%
  \institution{Kuaishou Technology}
  \city{Beijing}
  \country{China}}
\email{wangxueliang03@kuaishou.com}

\author{Qingpeng Cai$^\ddagger$}
\affiliation{%
  \institution{Kuaishou Technology}
  \city{Beijing}
  \country{China}}
\email{caiqingpeng@kuaishou.com}

\author{Lantao Hu}
\affiliation{%
  \institution{Kuaishou Technology}
  \city{Beijing}
  \country{China}}
\email{hulantao@kuaishou.com}

 \author{Han Li}
 \affiliation{%
   \institution{Kuaishou Technology}
   \city{Beijing}
   \country{China}}
 \email{lihan08@kuaishou.com}

\author{Peng Jiang$^\ddagger$}
\affiliation{%
  \institution{Kuaishou Technology}
  \city{Beijing}
  \country{China}}
\email{jiangpeng@kuaishou.com}

\author{Kun Gai}
\affiliation{%
 \institution{Unaffiliated}
 \city{Beijing}
 \country{China}}
\email{gai.kun@qq.com}

\author{Guangming Xie $^\ddagger$}
\thanks{$^\ddagger$ Corresponding author}
\affiliation{%
  \institution{Peking University}
  \city{Beijing}
  \country{China}}
\email{xiegming@pku.edu.cn}

\renewcommand{\shortauthors}{Xiaobei Wang et al.}


\begin{abstract}
In recommender systems, reinforcement learning solutions have shown promising results in optimizing the interaction sequence between users and the system over the long-term performance. 
For practical reasons, the policy's actions are typically designed as recommending a list of items to handle users' frequent and continuous browsing requests more efficiently. 
In this list-wise recommendation scenario, the user state is updated upon every request in the corresponding MDP formulation. 
However, this request-level formulation is essentially inconsistent with the user's item-level behavior.
In this study, we demonstrate that an item-level optimization approach can better utilize item characteristics and optimize the policy's performance even under the request-level MDP.
We support this claim by comparing the performance of standard request-level methods with the proposed item-level actor-critic framework in both simulation and online experiments.
Furthermore, we show that a reward-based future decomposition strategy can better express the item-wise future impact and improve the recommendation accuracy in the long term.
To achieve a more thorough understanding of the decomposition strategy, we propose a model-based re-weighting framework with adversarial learning that further boost the performance and investigate its correlation with the reward-based strategy.
\end{abstract}






 \maketitle

\section{Introduction}\label{sec: introduction}
Recommender systems serve as the user interfaces for a wide range of web services including e-commerce, news, and short-video platforms, and the general purpose is to filter contents of interests for users and increase their engagement/interaction with the system.
Recently, the optimization of users' engagement through reinforcement learning~\cite{afsar2022reinforcement, xin2022rethinking, cai2023reinforcing, cai2023two, xue2022resact, xue2023prefrec} has become a subject of significant interest, since it has the ability to optimize not only the user's immediate feedback on the recommended items but also the long-term rewards of future interactions.
The key idea is to model the user-system interaction~\cite{zhao2021usersim, gao2023alleviating} sequence as a Markov Decision Process (MDP)~\cite{mnih2015human,schulman2017proximal,pan2020softmax}.
Specifically, each context-aware request is represented by a user state that dynamically changes after each interaction, and the policy infers a recommendation action based on the current request.
Then, the user provides immediate feedback for the recommendation, which is later used to calculate the action's reward during training.
Adopting the general reinforcement learning (RL) paradigm, the goal is to maximize a cumulative reward over the expected future interactions.

For practical reasons, the recommendation policy is typically designed to provide a list of items/contents rather than a single item as an action in order to efficiently process the frequent and continuous browsing requests~\cite{ie2019slateq}.
This design choice is often seen in scenarios like short video, news, and blog recommendation systems, where users may typically consume/browse a large number of contents in a short time.
However, this list-wise recommendation mechanism is fundamentally inconsistent with the user's item-wise behavior.
Intuitively, a user only pays attention to one item at a time and his/her actual ``state'' changes after the interaction with each item, but the system is assumed to infer a list, obtain a list-wise feedback, and then update the user's state on each request.

In this paper, we emphasize this inconsistency in RL-based recommendation and formulate a request-level MDP with item-wise rewards.
Specifically, the system can only observe how the state changes in the request-level MDP, but we assume that the ground-truth item-level state transition in the user's viewpoint is not observable. 
A straightforward solution under the request-level MDP would simply optimize the cumulative list-wise reward using any standard RL approach~\cite{mnih2016asynchronous,timothy2016ddpg}, but this may induce information loss on the item characteristics for future reward estimation.
As an intuitive example, Tom receives a list of movie recommendations from either sports or documentary categories and gives positive feedback to sports movies but no feedback to documentary ones. 
If using a list-level optimization, the system may not discover the item-level differences and Tom's preference for sports. 
Fortunately, the item-wise user feedback throughout the interaction history is available, which augments the list-wise reward with a more detailed view.
Yet, each items' long-term value consists of both the immediate feedback and its future impact.
And it is reasonable to believe that the causal effects of a certain item on the user's future interactions\cite{gao2023alleviating} are not evenly distributed in the list of exposure.
Following the previous example, recommending sports movies may satisfy Tom's demands and gain more trust from him, so potentially increasing Tom's engagement in the future.
In contrast, recommending a documentary might be an unsatisfactory action as it could spoil the system's impression to Tom and lead to a decrease in user engagement in the future.
In general, we need an RL-based solution that can optimize the policy under the list-wise state transition, but also exploits both the item-level reward and item-wise future impact.

\begin{figure*}[ht]
\centering
\includegraphics[width=\linewidth]{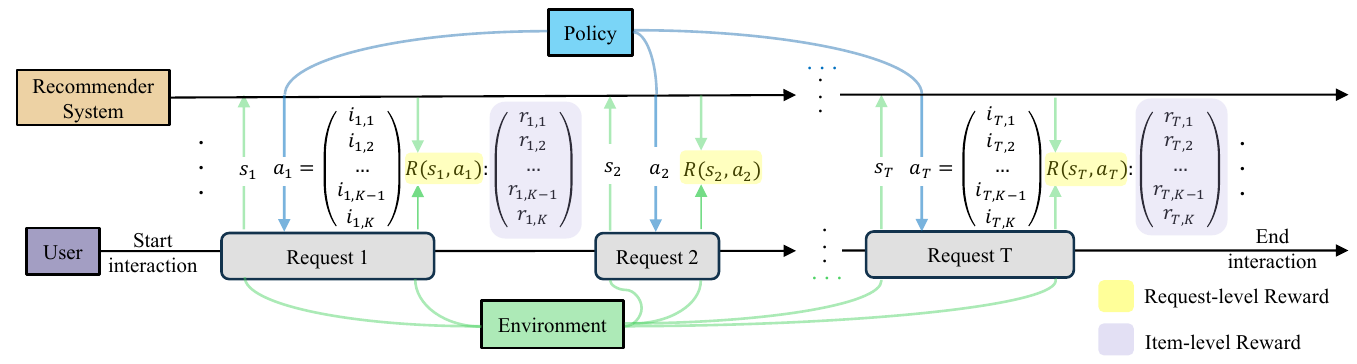}
\caption{Request-level MDP with observable item-level reward. A special case is K=1 that is equivalent to item-level MDP.}\label{fig: list_item}
\end{figure*}


To verify the aforementioned assumptions, we first propose an item-decomposed advantage actor-critic (ItemA2C) learning framework to distinguish items with different immediate feedback.
The model optimizes the same cumulative reward in the request-level MDP as standard request-level A2C, but both the critic's temporal difference loss and the actor's advantage maximization loss are decomposed into an item-wise optimization.
The empirical results demonstrate that this item-level paradigm is superior to its request-level counterpart. 
To further investigate how the item decomposition mechanism manipulates the optimization of the future impact of items, we design a re-weighting strategy where the evaluation of each item's future value may vary with respect to its immediate reward.
In other words, this strategy assumes that better immediate rewards indicate better future impact.
The resulting framework theoretically recovers the request-level actor-critic framework while providing separate item-wise long-term value estimation.
Knowing that there may exist other valid choices in the re-weighting space, we generalize the heuristic design of weights into a weight model which is optimized through adversarial optimization.

We summarize our contribution as follows:
\begin{itemize}
    \item We specify the challenge of inconsistency between users' item-wise view and the recommender system's list-wise view, and formulate a more practical request-level MDP formulation with item-wise reward.
    \item We propose an item-decomposed advantage actor-critic (Item-A2C) framework and verify its superiority on multiple public datasets and an online A/B test.
    \item We further illustrate and validate the correctness of the item-level future impact decomposition by a re-weighting strategy under the Item-A2C framework. And we theoretically show that the decomposition still recovers the request-level A2C in the objective functions.
    \item We also extend the heuristic approach into a learnable weight model that can further improve the recommendation performance, proving the existence of a wider feasible re-weighting space.
\end{itemize}
In the remainder of this paper, we first illustrate the challenges in RL-based recommendations and existing related solutions in section \ref{sec: related_work}.
Then section \ref{sec: method} presents our solution framework with theoretical analysis, and section \ref{sec: experiments} describes the experiments that address our claims.

\section{Related Work}\label{sec: related_work} 

\subsection{List-wise Recommendation}

For recommendation scenarios that take a list of items as an action, they can generally categorized in to one-step list recommendation and sequential list-wise recommendation problems.
The first study mainly focuses on the items mutual influences~\cite{cao2007learning,xia2008listwise,burges2010ranknet,ai2018learning}, and early approaches only discuss this issue in a top-k recommendation scenario where the goal is to directly optimize a list-wise metric.
The latter research efforts provide solutions to a more complicated sequential next-list recommendation scenario, where the list-wise MDP is introduced~\cite{rendle2010factorizing,ie2019slateq,liu2023generative}.
As one of the most representative works, SlateQ~\cite{ie2019slateq} is a reinforcement learning method for recommender systems (RL4RS).
It uses a DQN framework and proposes a decomposition technique that learns to distribute the list's Q-value toward each individual item,
and formulates the optimization problem of finding a slate with the maximum expected Q-value as a fractional linear program. 
Though SlateQ is effective in modeling the list-wise reward, its "single-choice" assumption mismatches our setting where all items in the lists may have non-zero rewards.
In general, we distinguish our problem setting from existing works as it emphasizes the existing item-wise feedback under the sequential list-wise recommendation scenario.

\subsection{Reinforcement Learning for Recommendation}

The RL-based recommendation system(RS)~\cite{afsar2021reinforcement,sutton2018reinforcement,zhao2018deep} adopts the MDP framework and aims to optimize the cumulative reward that reflects long-term user satisfaction. 
Although tabular-based~\cite{mahmood2007learning} methods can be used to optimize an evaluation table in a simple setting, they are limited to a fixed set of state-action pairs. 
Therefore, value-based methods~\cite{pei2019value,taghipour2007usage,zhao2021dear,zhao2018recommendations} and policy gradient methods~\cite{chen2019large, chen2019top, ge2021towards,ge2022toward, li2022autolossgen,xian2019reinforcement} have been proposed to learn to evaluate and optimize the action policy based on the sampled long-term reward. 
The actor-critic~\cite{zhao2018deep, xin2020self, zhao2020whole,cai2023two} paradigm integrates these two methods by simultaneously learning an action evaluator and an action generator. 
Our method belongs to this later paradigm.
Despite its effectiveness, reinforcement learning also faces several challenges when accommodating the recommender system, including but not limited to the exploration of combinatorial state/actions space~\cite{dulac2015deep,ie2019slateq,liu2020state, liu2023exploration}, unstable user behavior~\cite{bai2019model,chen2021user}, heterogeneous user feedback~\cite{chen2021generative,cai2023two}, and multi-task learning~\cite{meng2023hierarchical,liu2024sequential,liu2023multi}.
Among all existing works, we consider HAC~\cite{liu2023exploration} as closest to our work since it also included item-wise components in an RL framework.
In comparison, HAC aims to represent a list in a hyper-embedding space for end-to-end training, while our work focus on the item-level long-term effect modeling.
Note that we also emphasize the importance of item-level future impact attribution which could compliment the HAC model.

\section{Method}\label{sec: method}

\begin{figure*}[ht]
\centering
\includegraphics[width=\linewidth]{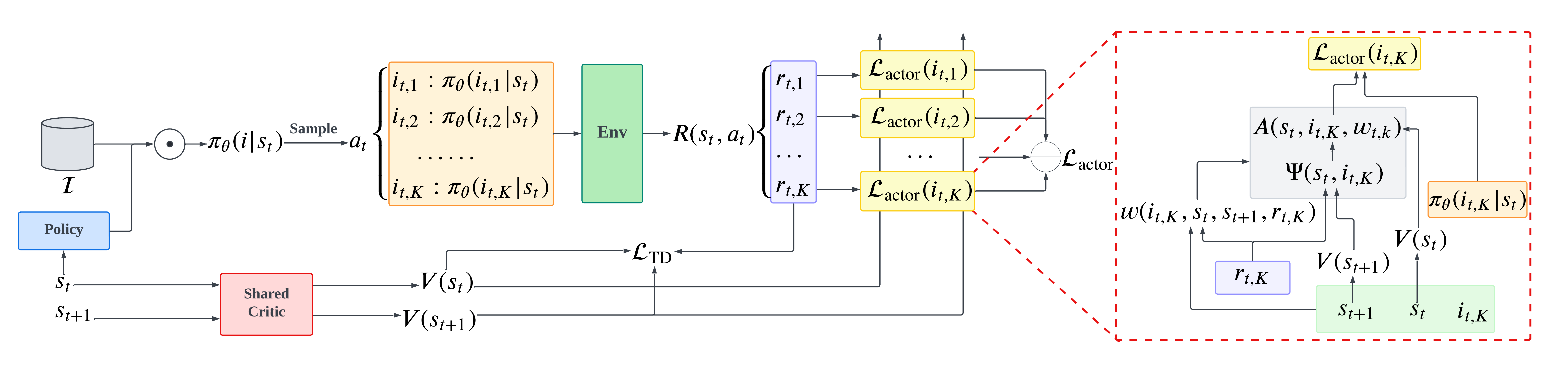}
\caption{The ItemA2C learning framework, where $\odot$ repesents the score function items and $\oplus$ represents summation.}\label{fig: Framework}
\end{figure*}

In this section, we first explain the request-level MDP formulation and illustrate our solution framework that consists of an actor-critic learning backbone, an item-wise decomposition framework, and two reweighting methods for item-wise future impact.
As an overview, Figure \ref{fig: list_item} provides the details of the problem formulation and Figure \ref{fig: Framework} summarizes our learning framework.

\subsection{Problem Formulation}\label{sec: method_problem_formulation}

In this work, we formulate the recommendation problem as a request-level MDP with item-wise rewards and provide an overview as Figure \ref{fig: list_item}.
Specifically, the system considers an item candidate pool $\mathcal{I}$ of size $N$, 
and we consider a recommendation policy that is responsible for providing an item list of size $K$ to the user upon each user request.
The user interacts/consumes the recommended items in the list one by one (e.g., watching each short video) and provides personalized feedback(e.g. click/like) that will later be used to calculate the rewards.
Then we can express the components of the MDP as follows:
\begin{itemize}
\item  $\mathcal{S}$: the continuous representation space of user state. Note that a user state encodes the user request which contains information of the user's static profile features (e.g. gender and age) as well as recent interaction history.
\item  $\mathcal{A}$: the action space corresponds to the possible recommendation lists. Without loss of generality, we consider the list of fixed size $K$ so the action space is $\mathcal{A} = \mathcal{I}^K$. And we can define each list-wise action as $a_t=[i_{t,1},\dots,i_{t,K}]$.
\item  $R(s_t,a_t)$: the immediate reward observed after taking action $a_t \in \mathcal{A}$ on state $s_t \in \mathcal{S}$. In our setting, we assume that this list-wise reward is a linear aggregation of item-wise reward, i.e. $R(s_t,a_t) = \sum_{i_{t,k}\in a_t} r(s_t,i_{t,k})$ where each item-wise reward $r_{t,k}=r(s_t,i_{t,k})$ is also observable.
\item  $\mathcal{P}$ : transition probability $P(s_{t+1}|s_t,a_t)$ reflects the probability reaching the next state $s_{t+1}$ from the current state $s_t$ given certain action $a_t$. We assume that this is implicitly modeled by the user request encoder as it takes user interaction history as input. As we have discussed in section \ref{sec: introduction}, the state change can only be seen at this request level and the item-wise state transition is not observable.
\item $\pi:\mathcal{S} \rightarrow \mathcal{A}$: the recommendation policy that chooses a list of items at each request.
\end{itemize}
And given this request-level MDP with item-wise feedback, the sample records in the experience replay buffer take the form \\
$(s_t, a_t, r_{t,1},\dots,r_{t,K}, s_{t+1}, d)$ where $d\in\{0,1\}$ represents whether the session ends after taking action $a_t$.

In summary, we present three distinguishable assumptions in our formulation:
1) the request-level state transition; 2) the observable item-wise reward; and 3) list-wise reward as a linear aggregation of item-wise reward.
Our \textbf{Goal} is to learn a policy that maximizes the user's expected cumulative reward over the interactions:
\begin{equation}
    \mathbb{E}[\mathcal{R}(s_t,a_t)] = \mathbb{E}[\sum_{i=0}^\infty \gamma^i R(s_{t+i},a_{t+i})]\label{eq: cumulative_reward}
\end{equation}

\subsection{Request-level A2C}\label{sec: method_backbone}

To better investigate the differences between item-wise and list-wise optimization, we first develop an advantage actor-critic (A2C) learning algorithm as the backbone that follows a standard RL framework~\cite{mnih2016asynchronous}.
We denote this framework as ``request-level A2C''.
It consists of an actor $\pi_\theta (a_t|s_t)$ that serves as the recommendation policy, and a critic $V(s_t)$ that aims to model the expected long-term cumulative reward of a given state (as in Eq.\eqref{eq: cumulative_reward}) and guide the optimization of the actor.

During optimization, the training samples come from an experience replay buffer and each sample takes the form of $(s_t,a_t,R(s_t,a_t)$ $,s_{t+1},d)$.
The optimization of critic networks follows a time-difference (TD) minimization objective:
\begin{equation}
\begin{aligned}
\mathcal{L}_\mathrm{critic} &= \Big(\Psi(s_t,a_t) - V(s_t)\Big)^2\\
\Psi(s_t,a_t) &=R(s_t,a_t) + \gamma (1-d) V(s_{t+1})\label{eq: listwise_critic_loss}
\end{aligned}
\end{equation}
where $\Psi$ represents the target value for the current state's critic estimation and $V(s_{t+1})$ is the target critic network that is widely used in RL to regulate the optimization of $V(s_{t})$ and stabilize the training.
When the critic is well-trained, it can guide the learning of the actor through the following advantage-boosting objective:
\begin{equation}
\begin{aligned}
    \mathcal{L}_\mathrm{actor} & = - A(s_t,a_t) \log \pi_\theta(a_t|s_t)\label{eq: listwise_actor_loss}\\
    A(s_t,a_t) & = R(s_t,a_t) + \gamma (1-d) V(s_{t+1}) - V(s_t)
\end{aligned}
\end{equation}
where the advantage estimates how much a certain action $a_t$ results in a better state than the average.
Note that we can also regard this advantage score as the difference between the target state-action value and the expected value of the current state, i.e. $A(s_t,a_t) = \Psi(s_t,a_t) - V(s_t)$.
This indicates that an action that generates a higher-than-average target value will tend to increase its recommendation probability.

\subsection{Item-wise Decomposition of A2C}\label{sec: item_decomposition}

The aforementioned request-level A2C framework is a pure list-wise approach that does not fully exploit the item-wise reward, so we introduce our decomposition method under the setting of section \ref{sec: method_problem_formulation}.
Specifically, we take advantage of the linear relationship between the item-wise reward and list-wise reward (i.e. $R(s_t, a_t) = \sum_{i_{t,k}\in a_t}r_{t,k}$) 
and derive an item-wise target function credit assignment for each item $i_{t,k}\in a_t$:
\begin{equation}
    \Psi(s_t,i_{t,k}) = r_{t,k} + \frac{1}{K} \gamma(1-d) V(s_{t+1})\label{eq: item_target_value}
\end{equation}
where the same state is shared for all items in the same list due to the unavailable item-wise state transition under the request-level MDP.
In this decomposition, there is no need to break the list-wise state transition, and each state value $V(s_t)$ is reused for K times when calculating the target for items in the list.


\subsubsection{Weighting Item-wise Future Impact}

In general, we believe that the total effect of an item $i_{t,k}$ in the list $a_t$ includes both the short-term and long-term perspectives.
While the immediate reward $r_{i,k}$ expresses the short-term effect, the items' share/split of the future impact $V(s_{t+1})$ reflects their long-term effect.
The direct decomposition in Eq.\eqref{eq: item_target_value} can distinguish items through their immediate reward but not the future impact because of the evenly distributed $V(s_{t+1})$.
Following this limitation, a natural question is whether we can break the uniform split.
And if we can, how should we address the future influence?

Intuitively, one possible assumption is that the items with better immediate user feedback are expected to impress the user more positively and thus improve the user's future behaviors, as we have discussed with examples in section \ref{sec: introduction}.
To verify this assumption, we propose a future re-weighting strategy that assigns $V(s_{t})$ in the target function with a weight that is positively related to the item's immediate reward:
\begin{equation}
\begin{aligned}
\Psi_{w}(s_t,i_{t,k}) &= r_{t,k} + w_{t,k}( \gamma (1-d)V(s_{t+1}))\\
w_{t,k} &= \frac{\alpha r_{t,k} + (1-\alpha)}{\alpha R(s_t,a_t) + (1-\alpha)K}\label{eq: weighted_item_target_value}
\end{aligned}
\end{equation}
where $\alpha$ is the hyperparameter that controls the balance between the pure re-weighting strategy and pure equal-weight strategy.
As two special cases in this design: when setting $\alpha=0$, these weights recover the equal-weight strategy Eq.\eqref{eq: item_target_value}; when setting $\alpha=1.0$, these weights become a pure reward-based strategy $w_{t,k} = r_{t,k} / R(s_t,a_t)$.
Note that in practice, the weight should always be made positive for consistent learning behavior, but we are NOT restricting $\alpha$ to $[0,1]$, since setting $\alpha<0$ means that immediate reward is assumed negatively related to the future impact and setting $\alpha>1$ indicates a negative bias on the immediate reward's relation to the future impact.
In section \ref{sec: experiments}, we will illustrate the effect of $\alpha$ in this design with empirical results.

For actor training, rather than $\pi_\theta (a_t|s_t)$, we can apply the item-wise optimization on $\pi_\theta (i_{t,k}|s_t)$ with the modified actor loss:
\begin{equation}
\begin{aligned}
    &\mathcal{L}_\mathrm{actor}(i_{t,k})  = - A(s_t,i_{t,k},w_{t,k}) \log \pi_\theta(i_{t,k}|s_t)\label{eq: itemwise_actor_loss}\\
    &  A(s_t,i_{t,k},w_{t,k}) = \Psi_w(s_t,i_{t,k}) - \frac{V(s_t)}{K}
\end{aligned}
\end{equation}
where the advantage estimates how much a certain item affects the cumulative reward compared to average items.


\subsubsection{Model-based Future Impact Re-weighting}

To further verify the idea of future impact re-weighting, explore the reweighting space, and analyze the correctness of the reward-based re-weighting strategy in the previous section, we also design a model-based solution that can automatically learn to reweigh the future impact decomposition.
Specifically, we generalize the original reward-based weighting strategy into a neural model $w_{t,k}=w(i_{t,k}, s_t, s_{t+1}, r_{t,k})$ which takes not only the item's immediate reward but also the future state and the corresponding item features as input.
These extra inputs are included because both the items feature and the next state $s_{t+1}$ are relevant to the item's future impact.
The model's output corresponds to the respective weights that replace $w_{t,k}$ in Eq.\ref{eq: weighted_item_target_value} and we ensure that the sum of weights equals one by applying a softmax function across the items in the same list.

One challenge for including this weight model is the lack of clear learning signal.
In practice, we design an adversarial learning framework and find it effective with the following objective:
\begin{equation}
\begin{aligned}
    \mathcal{L}_{weight} &= - \sum_{i_{t,k}\in a_t} \mathcal{L}_\text{actor}(i_{t,k}) \\ &= \sum_{i_{t,k}\in a_t} A(s_t,i_{t,k},w_{t,k}) \log \pi_\theta(i_{t,k}|s_t)\label{eq: weight_model_loss}
\end{aligned}
\end{equation}
which is also an item-level optimization.
This objective in Eq.\eqref{eq: weight_model_loss} is essentially the opposite of the actor loss Eq.\eqref{eq: itemwise_actor_loss}.
Intuitively, this will make the actor ``harder'' and force the actor to learn from ``harder'' samples.
Yet, we notice that the actual learning behaviors of the weight model and the actors could be far more complicated as we will describe in section \ref{sec: experiments}.
Note that there might also exist other valid solutions like hyperparameter searching~\cite{feurer2019hyperparameter} or auto machine learning approaches~\cite{fang2020rethinking}, we consider them as complementary to our research and focus on proving the existence of better weighting models than the reward-based strategy.


\subsection{Overall Item-level Learning Framework}\label{sec: item_learning_framework}

\begin{algorithm}[t]
\caption{ItemA2C Learning Framework}
\begin{algorithmic}[1]
\Procedure{Online Training}{}{}
\State Initialize all trainable parameters in the actor and critics.
\State Initialize the weight model if using model-based reweighting.
\While{True}
    \State Obtain mini-batch sample from the replay buffer.
    \State Mini-batch gradient descent for $V(s_t)$ with Eq.\eqref{eq: listwise_critic_loss}.
    \State Soft update of target network $V(s_{t+1})$.
    \State Mini-batch gradient descent for $w(\cdot)$ with Eq.\eqref{eq: weight_model_loss} if using model-based reweighting.
    \State Mini-batch gradient descent for $\pi$ with Eq.\eqref{eq: itemwise_actor_loss} for $i_{t,k}\in a_t$.
\EndWhile
\EndProcedure
\end{algorithmic}\label{alg: itema2c}
\end{algorithm}

In general, the overall learning framework involves a TD learning for critics as Eq.\eqref{eq: listwise_critic_loss}, an advantage-based actor learning (with or without reweighting) as Eq.\eqref{eq: itemwise_actor_loss}, and an adversarial learning for the weight model as Eq.\eqref{eq: weight_model_loss}.
Algorithm \ref{alg: itema2c} summarizes the details of the step-wise training and Figure \ref{fig: Framework} summarizes the learning of the decomposed A2C with model-based reweighting.
During the weight model optimization, the actor network and critic networks are fixed.
Similarly, the weight model is fixed during the critic's TD learning and the actor learning.

Mathematically, the weights of the future impact have property $\sum_{i_{t,k}\in a_t} w_{t,k} = 1$, which means the item-wise target function can reconstruct the original list-wise target:
\begin{equation}
\begin{aligned}
    \sum_{i_{t,k} \in a_t}\Psi_w(s_t,i_{t,k}) & = \sum_{i_{t,k}\in a_t} r_{t,k} + \sum_{i_{t,k}\in a_t} w_{t,k}\gamma(1-d) V(s_{t+1})\\
    & = R(s_t, a_t) + \gamma(1-d)V(s_{t+1}) = \Psi(s_t,a_t)
\end{aligned}
\end{equation}
which means that we can safely use the value functions $V(s_{t})$ and $V(s_{t+1})$ when guiding the actor learning, since the TD minimization is the same as Eq.\eqref{eq: listwise_critic_loss} which only uses the request-level target function $\Psi(s_t,a_t)$ and ignores the details in the item-wise target functions $\Psi(s_t,i_{t,k})$.
Though it is possible to apply a similar re-weighting strategy in the TD learning, we believe keeping the request-level critic learning is still a better choice for the items' mutual interactions within the same list can only be modeled in the request-level.
In other words, we only apply the item decomposition and reweighting during item-wise actor learning.

\section{Experiments}\label{sec: experiments}

To verify the effectiveness of the item-level decomposition under request-level MDP and our proposed reweighting methods, we first empirically study their performance and learning behavior in an online simulator of two public datasets as pre-online evaluation, then further validate our solution in online A/B test.

\subsection{Offline Experiments with Simulator}\label{sec: experiments_offline}

\begin{table*}[ht]\centering
    \caption{The full outputs for main results. The best performances in bold and second best in Underline.}
  \centering
  \begin{tabular}{cccccccc}
    \toprule
    \multirow{3}{*}{type} & \multirow{3}{*}{Model} & \multicolumn{6}{c}{ML1M} \\
    \cmidrule{3-8}
   & & \multicolumn{3}{c}{Total Reward} & \multicolumn{3}{c}{Depth}\\
   \cmidrule(r){3-5}\cmidrule(l){6-8}
   & &  Average & Maximum & Minimum & Average  & Maximum & Minimum \\
   \toprule
      \multirow{3}{*}{request-level}
    & DDPG & 12.97 $ \pm(2.95)$   & 19.47 $ \pm(0.76)$ & 2.52 $ \pm(1.37)$ & 13.72  $ \pm(2.62)$ & 19.52 $ \pm(0.69)$ & 4.49 $ \pm(1.21)$ \\
     & A2C  & 16.88 $ \pm(0.58)$   & 19.94 $ \pm(0.08)$ & 7.37 $ \pm(0.97)$ & 17.32  $ \pm(0.50)$ & 19.95 $ \pm(0.07)$ & 9.29 $ \pm(0.79)$ \\
     & HAC  & 17.53 $ \pm(0.13)$ & 19.97 $ \pm(0.03)$ & 7.78 $ \pm(1.26)$ & 17.90  $ \pm(0.12)$ & 19.97 $ \pm(0.03)$ & 9.67 $ \pm(1.10)$ \\

          \cmidrule{1-8}
&Supervision & 12.57 $ \pm(2.14)$ & 19.72 $ \pm(0.37)$ & 1.18 $ \pm(0.79)$ & 13.71 $ \pm(1.81)$ & 19.76 $ \pm(0.32)$ & 4.14 $ \pm(0.62)$\\

     \multirow{6}{*}{item-level}
   & SlateQ &4.62 $ \pm(0.19)$  & 12.04 $ \pm(0.41)$ & 0.06 $ \pm(0.07)$ & 6.96 $ \pm(0.16)$ & 13.19 $ \pm(0.35)$  & 3.22 $ \pm(0.04)$\\
      & itemA2C($\alpha=0$) &17.58 $ \pm(0.62)$  & 19.92 $ \pm(0.03)$ & 8.58 $ \pm(1.95)$ & 17.94 $ \pm(0.53)$ & 19.93 $ \pm(0.02)$ & 10.33 $ \pm(1.63)$ \\
     & itemA2C-W($\alpha=0.5$) &17.57 $ \pm(0.54)$  & 19.96 $ \pm(0.05)$ & 8.55 $ \pm(0.83)$ & 17.93 $ \pm(0.46)$ & 19.96 $ \pm(0.04)$ & 10.31 $ \pm(0.71)$ \\
     & itemA2C-W($\alpha=1$) & \underline{17.80 $ \pm(0.73)$} & \underline{19.97 $ \pm(0.06)$} & \underline{8.83 $ \pm(1.30)$} & \underline{18.12 $ \pm(0.63)$} & \underline{19.97 $ \pm(0.05)$} & \underline{10.81 $ \pm(1.12)$} \\
     & itemA2C-M &\textbf{17.94$ \pm(0.47)$}  & \textbf{20.00 $ \pm(0.02)$} & \textbf{9.79 $ \pm(1.22)$} & \textbf{18.24 $ \pm(0.40)$} & \textbf{20.00 $ \pm(0.02)$} & \textbf{11.33 $ \pm(1.02)$} \\
    \toprule
    \multirow{3}{*}{type} & \multirow{3}{*}{Model} & \multicolumn{6}{c}{KuaiRand} \\
    \cmidrule{3-8}
   & & \multicolumn{3}{c}{Total Reward} & \multicolumn{3}{c}{Depth}\\
   \cmidrule(r){3-5}\cmidrule(l){6-8}
   & &  Average & Maximum & Minimum & Average  & Maximum & Minimum \\
   \toprule

    \multirow{3}{*}{request-level}
     & DDPG & 11.71  $ \pm(0.98)$ & 19.43$ \pm(0.35)$ & 0.40$ \pm(0.42)$ & 12.93 $ \pm(0.83)$ & 19.50$ \pm(0.32)$ & 3.52$ \pm(0.34)$ \\    
    & A2C & 10.01$ \pm(0.86)$  & 19.47 $ \pm(0.28)$& -0.24$ \pm(0.10)$ & 11.56$ \pm(0.72)$ & 19.54$ \pm(0.25)$ & 3.04$ \pm(0.09)$\\
       &HAC & 12.65$ \pm(0.26)$  & 19.58 $ \pm(0.24)$& 0.82$ \pm(0.43)$ & 13.72$ \pm(0.22)$ & 19.63 $ \pm(0.22)$ & 3.86 $ \pm(0.34)$\\
       
          \cmidrule{1-8}
&Supervision & 9.78$ \pm(3.15)$  & 18.76 $ \pm(1.34)$& 0.16$ \pm(0.63)$ & 11.32$ \pm(2.66)$ & 18.92$ \pm(1.18)$ & 3.35$ \pm(0.51)$\\
     
     \multirow{6}{*}{item-level}
   & SlateQ &3.39 $ \pm(0.20)$  & 12.65 $ \pm(0.23)$ & -0.3  $ \pm(0.05)$& 5.99 $ \pm(0.17)$ & 13.67 $ \pm(0.19)$  & 3.00 $ \pm(0.04)$\\
      & itemA2C($\alpha=0$) &13.45 $ \pm(0.08)$  & 19.58 $ \pm(0.26)$ & 1.01 $ \pm(0.46)$ & 14.41 $ \pm(0.07)$ & 19.64 $ \pm(0.22)$ & 4.01 $ \pm(0.37)$ \\
     & itemA2C-W($\alpha=0.5$) &14.77 $ \pm(0.41)$  & \textbf{19.69 $ \pm(0.20)$} & 2.26 $ \pm(0.72)$ & 15.52 $ \pm(0.35)$ & \textbf{19.73 $ \pm(0.18)$} & 5.03 $ \pm(0.58)$ \\ 
    & itemA2C-W($\alpha=1$) & \underline{15.48 $ \pm(0.61)$}  & \underline{19.65 $ \pm(0.24)$} & \underline{3.58 $ \pm(0.92)$} & \underline{16.13 $ \pm(0.53)$} & \underline{19.69 $ \pm(0.21)$} & \underline{6.11 $ \pm(0.77)$} \\
    & itemA2C-M &\textbf{16.03 $ \pm(0.53)$} & 19.60 $ \pm(0.24)$ & \textbf{5.21 $ \pm(0.68)$} & \textbf{16.59 $ \pm(0.45)$} & 19.64 $ \pm(0.21)$ & \textbf{7.48 $ \pm(0.55)$} \\
     
    \bottomrule

  \end{tabular}
  
  \label{tab:result}
\end{table*}

\subsubsection{Datasets and Online Simulator}

We use two publicly available datasets: ML1M\cite{harper2015movielens} and KuaiRand1K\cite{gao2022kuairand}. 
The ML1M dataset includes users' rating records of movies, while KuaiRand1K is a dataset that contains user interaction records with short videos. 
We preprocess both datasets by a similar process in~\cite{zhao2023kuaisim}, in which each record is organized chronologically and users/items with less than 10 interactions are removed (10-core filtering).


Our simulator is constructed based on the session-based simulator as in KuaiSim \cite{zhao2023kuaisim} for both datasets. 
We train a user interaction model to estimate the probability of a user's click based on their dynamic interaction history and static profile features.
During online RL the user simulator will sample immediate feedback according to this model and serve as the interactive environment.
For simplicity, we directly consider the click-or-not signal as the reward, where a click results in a reward of 1.0 and a missing click leads to a reward of -0.2 following the same setting in~\cite{liu2023exploration}.
Though the reward design is not the focus of this work, one can easily generalize our solution for other reward settings as we will show in section \ref{sec: experiments_live}.
For all environments, we limit the maximum episode depths to 20 and set the list size to $K=6$.
When a user's episode ends, a new user is randomly picked to fill in the blank.
And each episode step simultaneously runs the simulator for a batch of 64 users.

\subsubsection{Evaluation Protocol}

With the online simulator ready, we can apply pseudo-online training for RL models, and evaluate its performance in the last 100 episode steps (minimum 320 user episodes).
We observed that the majority of RL-based techniques achieved convergence and stabilization within 30,000 iterations. 
For long-term evaluation metrics, we report the \textbf{total reward} (without discount) and \textbf{depth} of user sessions/episodes. 
Higher values for these metrics indicate superior performance.

\subsubsection{Baselines}

To better compare the proposed methods with other feasible solutions under the request-level MDP, we implemented the following baselines:
\begin{itemize}
\item SlateQ~\cite{ie2019slateq}: a DQN method that learns to estimate the Q-value of a state-item pair, it also considers the item-decomposition under the list-wise scenario.
\item DDPG~\cite{timothy2016ddpg}: 
a request-level Deep DPG method that uses a vectorized hyper-action to represent a whole list for both the actors and critics, so that end-to-end training could be applied for actor training.
\item Supervision: the supervised learning method which included a transformer-based history and profile encoder like that in SASRec and a dot product item selector like HAC that directly optimizes the item-wise scoring function through binary cross entropy loss on item-wise labels.
\item HAC~\cite{liu2023exploration}: The state-of-the-art request-level RL method which is also an actor-critic framework that uses a vectorized hyper action to represent each item list. The optimization process includes TD learning for a request-level critic, Q value maximization for a request-level actor, action space regularization and item-wise supervision.
\item A2C(section \ref{sec: method_backbone}): The request-level Actor-Critic method.
\end{itemize}
We present the details of the specifications in Appendix \ref{ap: model_specification}.

\begin{table*}[ht]\centering
  \caption{Performance for different list size.}
  \centering
  \begin{tabular}{ccccccccc}

    \toprule
    \multirow{3}{*}{list size} & \multicolumn{8}{c}{ML1M} \\
    \cmidrule{2-9}
    & \multicolumn{2}{c}{HAC} & \multicolumn{2}{c}{itemA2C($\alpha=0$)} & \multicolumn{2}{c}{itemA2C-W($\alpha=0.5$)} & \multicolumn{2}{c}{ItemA2C-M}\\
    \cmidrule(r){2-3}\cmidrule(lr){4-5}\cmidrule(lr){6-7}\cmidrule(l){8-9}
      & Reward   & Depth & Reward  & Depth & Reward & Depth & Reward  & Depth \\
          \cmidrule{1-9} 

      1 &\textbf{19.30$ \pm(0.28)$}  & \textbf{19.61$ \pm(0.15)$} & 19.10$ \pm(0.19)$   & 19.51$ \pm(0.12)$ & 19.10$ \pm(0.19)$  & 19.51$ \pm(0.12)$ & 19.10$ \pm(0.19)$   & 19.51$ \pm(0.12)$ \\
      2 & 18.66$ \pm(0.29)$  & 19.07$ \pm(0.21)$ & 18.69$ \pm(0.35)$ & 19.09$ \pm(0.25)$ & 18.84$ \pm(0.18)$  & \textbf{19.20$ \pm(0.14)$} & \textbf{18.85$ \pm(0.46)$}   & 19.08$ \pm(0.32)$\\
      4 & 18.14$ \pm(0.51)$  & 18.51$ \pm(0.41)$ & 18.17$ \pm(0.41)$ & 18.54$ \pm(0.34)$ & 18.30$ \pm(0.54)$  & 18.63$ \pm(0.44)$ & \textbf{18.39$ \pm(0.45)$}   & \textbf{18.72$ \pm(0.35)$}\\
      8 & 17.11$ \pm(0.70)$  & 17.45$ \pm(0.61)$ & 17.19$ \pm(0.47)$ & 17.51$ \pm(0.42)$ & 17.21$ \pm(0.83)$  & 17.54$ \pm(0.73)$ & \textbf{17.33$ \pm(0.48)$}   & \textbf{17.64$ \pm(0.42)$}\\
      
      16 & 15.65$ \pm(0.74)$  & 15.96$ \pm(0.69)$ & 16.12$ \pm(0.96)$  & 16.39$ \pm(0.91)$ & 16.23$ \pm(0.91)$ & 16.43$ \pm(0.90)$ & \textbf{16.27$ \pm(0.99)$}   & \textbf{16.51$ \pm(0.93)$}\\
      32 & 14.72$ \pm(0.82)$  & 14.93$ \pm(0.78)$ & 14.79$ \pm(1.02)$  & 15.00$ \pm(0.98)$ & 14.58$ \pm(0.73)$ & 14.68$ \pm(0.67)$ & \textbf{14.85$ \pm(1.13)$}  & \textbf{15.05$ \pm(1.09)$}\\
      
    \toprule
    \multirow{3}{*}{list size} & \multicolumn{8}{c}{KuaiRand} \\
    \cmidrule{2-9}
    & \multicolumn{2}{c}{HAC} & \multicolumn{2}{c}{itemA2C($\alpha=0$)} & \multicolumn{2}{c}{itemA2C-W($\alpha=0.5$)} & \multicolumn{2}{c}{ItemA2C-M}\\
    \cmidrule(r){2-3}\cmidrule(lr){4-5}\cmidrule(lr){6-7}\cmidrule(l){8-9}
      & Reward   & Depth & Reward  & Depth & Reward & Depth & Reward  & Depth \\
          \cmidrule{1-9} 

      1 & 16.03$ \pm(0.39)$  & 17.98$ \pm(0.20)$ & \textbf{16.93$ \pm(0.40)$}  & \textbf{18.42$ \pm(0.23)$}& \textbf{16.93$ \pm(0.40)$}  & \textbf{18.42$ \pm(0.23)$} & \textbf{16.93$ \pm(0.40)$}  & \textbf{18.42$ \pm(0.23)$}\\
      2 & 15.01$ \pm(0.29)$  & 16.59$ \pm(0.19)$ & 15.50$ \pm(0.29)$  & 16.92$ \pm(0.20)$ & 15.97$ \pm(0.51)$ & 16.53$ \pm(0.34)$ & \textbf{16.52$ \pm(0.27)$}  & \textbf{17.73$ \pm(0.19)$}\\
      4 & 13.37$ \pm(0.33)$  & 14.69$ \pm(0.25)$ & 14.07$ \pm(0.30)$  & 15.26$ \pm(0.24)$ & 15.03$ \pm(0.50)$ & 15.58$ \pm(0.40)$ & \textbf{16.29$ \pm(0.32)$}  & \textbf{17.02$ \pm(0.26)$}\\
      8 & 12.14$ \pm(0.71)$  & 13.03$ \pm(0.63)$ & 12.67$ \pm(0.39)$  & 13.50$ \pm(0.35)$ & 14.04$ \pm(0.50)$ & 14.43$ \pm(0.45)$ & \textbf{15.05$ \pm(0.62)$}  & \textbf{15.59$ \pm(0.55)$}\\
      
      16 & 11.51$ \pm(0.35)$  & 11.94$ \pm(0.30)$ & 11.28$ \pm(0.57)$  & 11.93$ \pm(0.53)$ & 12.74$ \pm(0.26)$ & 13.37$ \pm(0.24)$ & \textbf{13.73$ \pm(0.77)$}  & \textbf{14.18$ \pm(0.71)$}\\
      32 & 10.37$ \pm(0.22)$  & 10.71$ \pm(0.19)$ &  9.71$ \pm(0.64)$  & 10.11$ \pm(0.61)$ & 11.29$ \pm(0.36)$ & 11.63$ \pm(0.33)$ & \textbf{11.39$ \pm(0.41)$}  & \textbf{11.71$ \pm(0.39)$}\\

    \bottomrule

  \end{tabular}
  \label{tab:list_szie}
\end{table*}

\subsubsection{Model Alternatives}

We denote our item-decomposition framework with equal weights as \textbf{itemA2C}, 
the reward-based re-weighting strategy as \textbf{itemA2C-W}, 
and the extended framework that uses weight model with adversarial learning is noted as \textbf{itemA2C-M}(model).
For reproduction of our empirical study, we provide implementation and training details as well as the hyperparameters of best results in our grid search space in our released source code~\footnote{https://github.com/wangxiaobei565/ItemDecomposition}.

\begin{figure}[t]
\centering
  \includegraphics[scale=0.245]{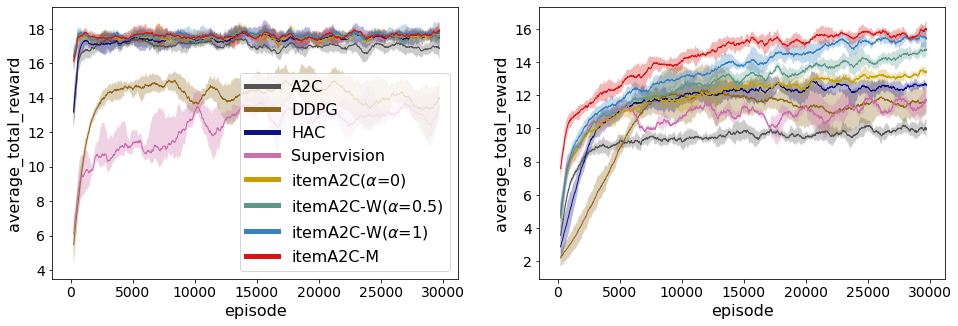}
\smallskip
\text{\ \ \ \ (a) \ Performance in ML1M  \ \ \ \ \ \ \  (b) \ Performance in KuaiRand}
\caption{Learning curves of all methods.}\label{fig: main_results}

\centering
\end{figure}

\begin{figure}[t]
\centering
  \includegraphics[scale=0.245]{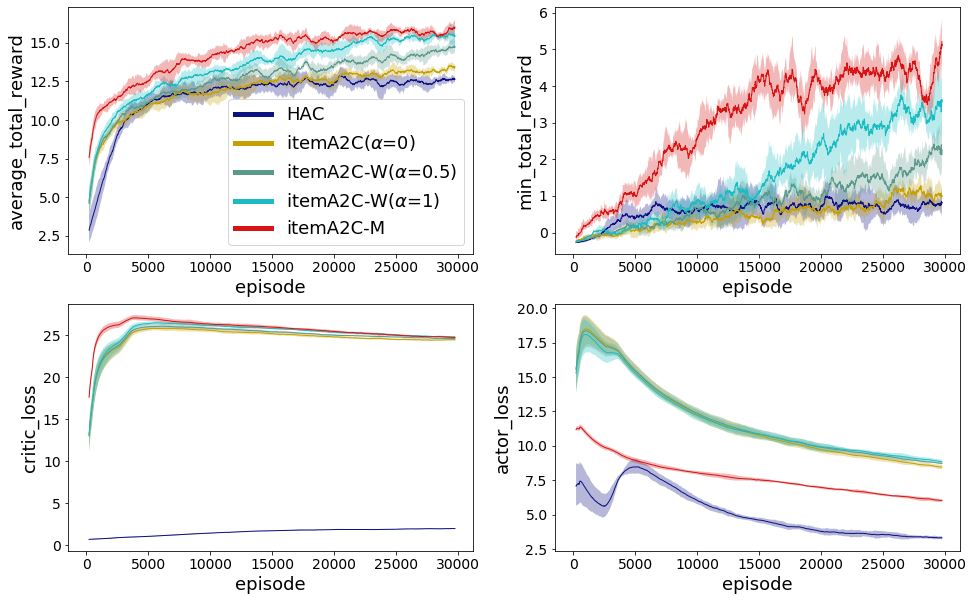}
\caption{Training curves of all methods on KuaiRand. We also show the loss curves about critic and actor. }\label{fig: learning_curves}
\centering
\end{figure}

\subsubsection{Main Results}\label{sec: experiments_offline_main_result}

For each experiment across all models, we run online training and evaluation for five rounds with different random seeds and report the average results in Table \ref{tab:result}.
We can see that most RL-based methods (DDPG, A2C, HAC, and ItemA2C) can improve the recommendation result over supervised learning, indicating the superiority of optimizing the cumulative reward over immediate reward.
The SlateQ method applies item decomposition but is not well suited for our problem setting since it decomposes the request-level total reward without using the item-wise reward, which results in sub-optimal performance.
Among the baselines, the best request-level method is HAC in terms of the total reward and depth metrics.
Including the itemwise supervision might be the reason why HAC surpasses other baselines, but its lack of item-wise decomposition of future impact is an inferior design which is reflected in the empirical results compared to our itemA2C alternatives.
In comparison, ItemA2C-M with weight model achieves the best overall performance and significantly (with $p<0.05$) improves the reward by 27\% and the depth by 20\% in KuaiRand1K and marginally (though not statistically significant) improves the reward by 2.3\% and depths by 1.8\% in ML1M.
Currently, we believe the difference between the improvement made on ML1M and KuaiRand is related to the consistency of user behaviors (more consistent interactions in the short video platform than that in the movie) and the number of candidate items (3706 items in ML1M and 11643 items in KuaiRand).
These data characteristics make the item selection process easier for the baseline models on ML1M dataset to approach the optimal performance.

Additionally, compared to the request-level A2C baseline, all of our ItemA2C alternatives demonstrated significant improvement in both user environments. 
Besides, All item-level RL methods have relatively the same level of reward variance and appear to be smaller than request-level methods, which can also be seen in their learning processes as shown in Figure \ref{fig: learning_curves}.
This verifies the effectiveness of the item-level decomposition and its consistent superiority and learning stability over its request-level counter-part.
In comparison to equal-weight future decomposition, both the strategy-based ItemA2C-W and model-based ItemA2C-M resulted in even better recommendation performance.
Moreover, increasing $\alpha$ in the strategy-based reweighting alternative amplifies future impact and further improves the recommendation result.
This proves the effectiveness of the future impact decomposition in Eq.\eqref{eq: weighted_item_target_value}.




\begin{figure}[t]
\centering
\includegraphics[width=\linewidth]{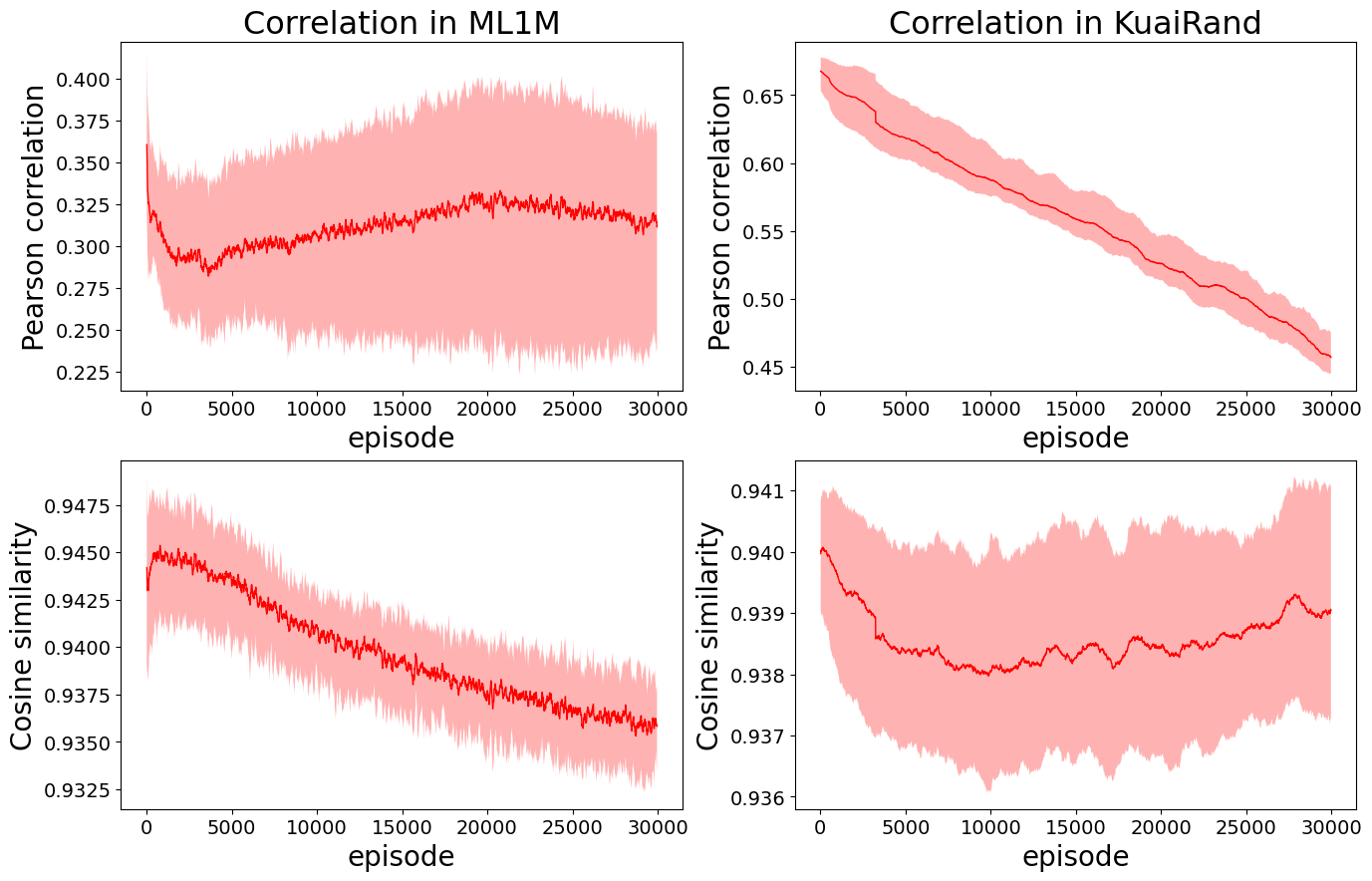}
\caption{The Trends of Cosine Similarity and Pearson Correlation Coefficient between weight model and heuristic re-weighting during training.}\label{fig: ad_train}
\end{figure}

\subsubsection{Strategic vs. Model-based Reweighting}
As shown in Table \ref{tab:result}, ItemA2C-M achieves better performance than ItemA2C-W in both long-term reward and reward variance.
This indicates the existence of other feasible solutions to the reweighting strategy for the item-level future impact decomposition.
To better understand the similarity and differences between the two methods, in Figure \ref{fig: ad_train} we plot the cosine similarity and the Pearson correlation coefficient between the weights of Eq.\eqref{eq: weighted_item_target_value} in ItemA2C-W and the learned weights in ItemA2C-M.
Our results reveal that the weights exhibit high cosine similarity and positive correlation between reward-based reweighting strategy and model-based extension in both datasets. 
This finding proves the validity of the design of Eq.\eqref{eq: weighted_item_target_value} assuming the generality of the model-based solution.
However, both metrics decrease at the beginning, which proves that the adversarial learning in model-based reweighting method may gradually find different feasible weighting strategies.
We also find inconsistent behaviours in that ItemA2C-M becomes increasingly dissimilar in the cosine metric and relatively stable in Pearson correlation in ML1M, but observes decreasing Pearson correlation and stable cosine similarity in KuaiRand.
Note that the Pearson correlation calculates item-wise correlation separately but cosine similarity calculates the relation between the entire weight vector of size $K$.
This means that the relative weights among items in ItemA2C-M tend to be more aligned with Eq.\eqref{eq: weighted_item_target_value} even though the two methods absolution values of weights diverge in KuaiRand.
In contrast, the relative weights in ItemA2C-M slowly diverge from Eq.\eqref{eq: weighted_item_target_value} in ML1M.
In general, this may indicate that the optimal reweighting strategy of different recommendation services may not take the same form.


\subsubsection{Ablation}
\begin{figure}[tbp]
\centering
  \includegraphics[scale=0.3]{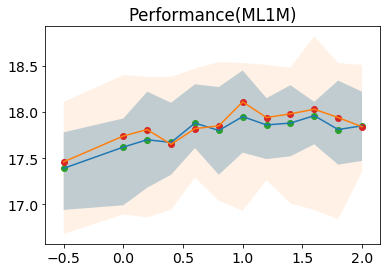}
  \hspace{0.001in}
  \includegraphics[scale=0.3]{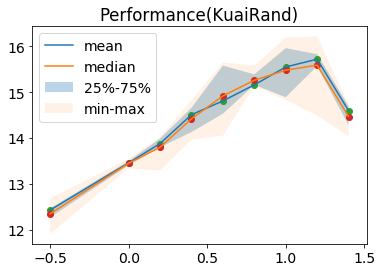}
\caption{Total reward under different $\alpha$ and  the results include mean, median, quartiles, and min-max.
}\label{fig: ablation}
\centering
\end{figure}

\textbf{Influence of $\alpha$:} Recall that $\alpha$ controls the balance between equal-weight strategy and full re-weighting strategy in ItemA2C-W.
In this study, we search the parameter space of $\alpha$ in [0.0, 2.0] and include a negative value $\alpha=0.5$ to represent a possible negative effect of reward-based re-weighting.
Note that $\alpha>1.0$ represents an ``over-weighting'' strategy where the equal-weight strategy contributes negatively to the value.
We present the resulting total session reward under different $\alpha$ in Figure \ref{fig: ablation}.
In both datasets, the best setting is observed between $[1.0,1.5]$, which further proves the effectiveness of the future impact reweighting.
Besides, the sensitivity of $\alpha$ is larger in KuaiRand than ML1M since the total reward in KuaiRand varies on a larger scale compared to its variance.
This is also related to the easier learning environment in the ML1M data than KuaiRand as discussed in section \ref{sec: experiments_offline_main_result}.

\textbf{Effect of list size: } 
The main experiment in our paper set the list size to $K=6$ to mirror the real scenario presented in the online system.
but it is reasonable to believe that the choice of list size also affects the recommendation result.
We conducted experiments with $K\in\{1,2,4,6,8,16,32\}$ which changes both the MDP of the RL model but also the user environment.
We test both request-level A2C and item-level ItemA2C, ItemA2C-W, ItemA2C-M and run each experiment for five rounds with different random seeds and report the average values in Table \ref{tab:list_szie}.
The performance of all methods gradually deteriorated as the list size increases.
This is a direct indicator of the existence of the inconsistent view between request-level system MDP and item-level user MDP, and the most accurate view aligns with the item-wise user view.
The proposed ItemA2C-M achieves the best result across different list size settings.
In our empirical study on both datasets, ItemA2C consistently achieves better performances than the best baseline HAC and all other methods for $K>2$.
Moreover, the superiority of ItemA2C-M over ItemA2C-W is also verified for different settings of $K$.

\subsection{Live Experiments}\label{sec: experiments_live}

\begin{table*}[t]
\caption{Online performances of item-level A2C and all results are statistically significant.}
\centering
\begin{tabular}{l|c|c|c|c|c}
\toprule
Method& Watch Time &  Like & Follow & Collect & Comment\\
\midrule
request-level A2C \textit{v.s.} baseline & -0.073$\%$ & +0.713$\%$ & +0.205$\%$ & +0.736$\%$ & +0.328$\%$  \\
ItemA2C \textit{v.s.} request-level A2C  & +0.129$\%$  & +1.103$\%$ & +0.300$\%$  & +0.963$\%$  & +0.221$\%$  \\
ItemA2C-W($\alpha=1$) \textit{v.s.} ItemA2C  & -0.013$\%$  & +0.451$\%$ & +0.636$\%$  & +0.616$\%$  & +0.258$\%$  \\
\bottomrule
\end{tabular}
\label{tab: online_results}
\end{table*}

\subsubsection{Experimental Setup} 
\begin{figure}[t]
\centering
\includegraphics[width=\linewidth]{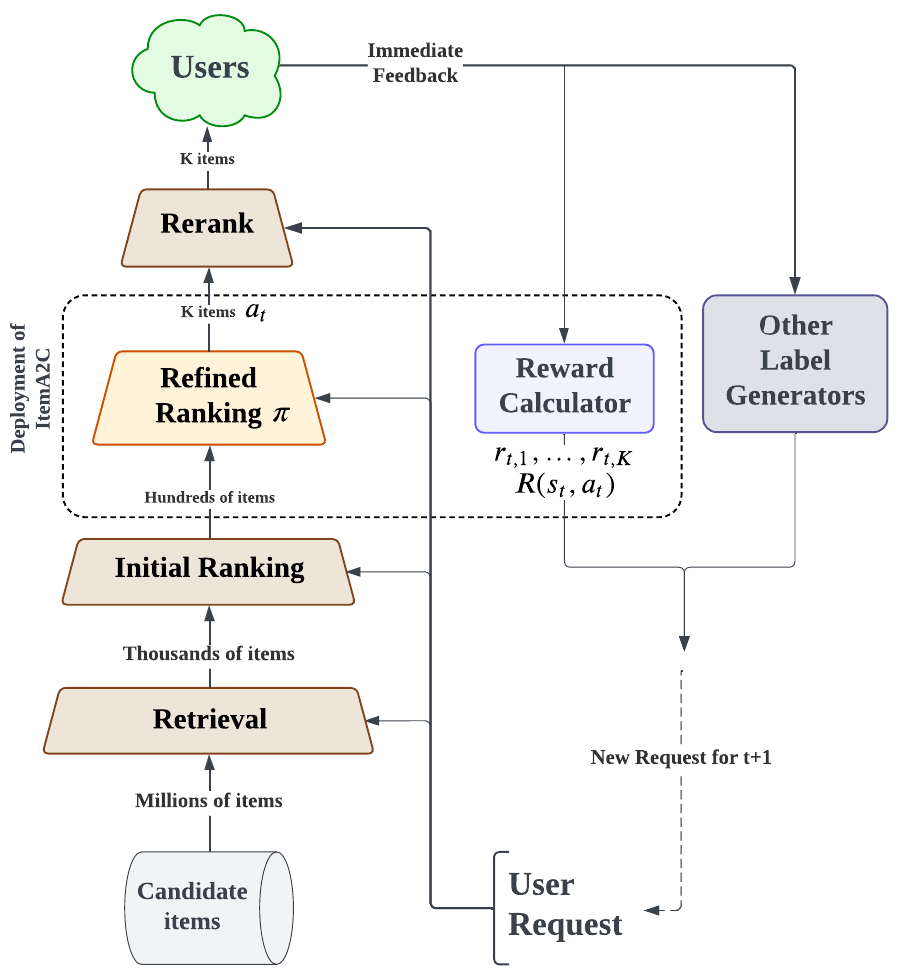}
\caption{The entire workflow consists of a relevant video retrieval stage, an efficient initial ranking stage, a refined ranking stage, and a final reranking . The ItemA2C is deployed in the refined ranking stage.}\label{fig: online_workflow}
\end{figure}

To further verify our proposed method in a real-world environment, we conduct A/B testing on an industrial video recommendation platform.
The platform serves over 100+ million users every day and Figure \ref{fig: online_workflow} summarizes the recommendation workflow.
Specifically, we are considering the recommendation task that recommends $K$ items for each user request, which corresponds to the refined ranking stage where the candidate set size is around 500 (filtered by previous retrieval and ranking stages) and the output list size is $K=6$. 
Different from the single behavior settings in our offline experiments, the user behaviors in the online system are more complicated and we include watch time, like, follow, collect, and comment as targets.
The corresponding item-wise reward design is a linear combination:
\begin{equation}
\begin{split}
    r(s_t,i_{t,k}) = w_1 * \mathbb{I}[\mathrm{quantile\_watch\_time}] + w_2 * \mathbb{I}[\mathrm{like}]\\
    + w_3 * \mathbb{I}[\mathrm{follow}] + w_4 * \mathbb{I}[\mathrm{collect}]
    + w_5 * \mathbb{I}[\mathrm{comment}]
\end{split}
\end{equation}
where the weight for each signal is set to a positive empirical value in $[0.1,1.0]$ and we use quantization for the watch time signal so that all feedback signals are in discrete space.

We randomly segregate the total traffic for the experiments, and 10\% of traffic for each of the selected methods including a request-level method (similar to A2C), ItemA2C, and ItemA2C-W.
For all three methods, the actor uses a 4-layer MLP as the network structure and the input encoding is a 400-dimension vector that encodes the user request and the candidate items.
The output ranking scores of items correspond to the policy output $\pi_\theta (i|s_t)$.
To give a fair comparison, we also include our baseline model (with 20\% traffic hold-out) that adopts learning-to-rank supervision at the item level.

\subsubsection{Results Analysis}

We keep each experiment online for one week and summarize the average results in Table \ref{tab: online_results}.
Note that the ``watch time'' metric expresses the average length the user watches a video without discretization, and all other metrics express the rate of a certain behavior on recommended videos.
Considering the overall performance over all metrics, the request-level A2C improves the (non-RL) baseline model on most metrics except for a slight decrease in watch time, which verifies the effectiveness of RL.
The ItemA2C can significantly improve the performance in all metrics over request-level RL, which validates our claim on the superiority of item-level optimization.
The ItemA2C-W further improves the metrics except an (statistically) insignificant decrease in watch time, which proves the effectiveness of future impact decomposition and that our proposed method can generalize to more complicated reward designs.
Additionally, we also observe that ItemA2C methods achieve a minor increase of daily active users (DAU) by 0.028\% and significant improvement in the user retention (indicating the probability of a user's return in the next week) by 0.016\%.
This means that the selected metrics are positively correlated with the user's activity level.

\section{Conclusion and Discussion}\label{sec: conclusion}

In this paper, we present the challenge of inconsistent viewpoints between request-level recommendation policy and item-level user behaviors. 
To address this challenge, we formulated a practical list-wise MDP formulation incorporating item-wise reward and proposed an item-wise decomposition method under the actor-critic learning framework.
Based on the proposed method, we show that one should not only utilize the item-wise reward for the optimization of each immediate user feedback but also distinguish each item's future impact through the value function decomposition.
The proposed method introduces no extra learnable parameters than A2C and is well-suited for item-wise parallel training.
We additionally proposed a model-based method with adversarial learning that automatically learns to decompose the future impact and proves that the reward-based strategy is reasonable but could be further improved.
Note that the model-based re-weighting solution may introduce extra modeling and training cost but no extra computation during inference. 
In terms of the data collection, one has to construct the data stream for the request-level MDP with item-wise feedback to implement the solution. 
Fortunately, this data is usually easy to collect in most industrial recommender systems. 
As an extension to our model-based approach, future research on new ways to represent the future impact of each item may also give insights into user behavior modeling, improving the recommendation systems and user satisfaction.

Additionally, our design does not restrict the reward settings to click-based reward design as in the offline experiment or linear combination of multi-response as in the online A/B test, future research on more complicated non-linear reward design (e.g. intra-list diversity) may further improve recommender performance.
During deployment, our approach is well-suited for the last ranking stage or the reranking stage where user feedback are provided for all the output items, and some extra efforts might needed to accommodate previous ranking stages, since the candidate pool size is much larger and the output size are not necessarily K.
In the RL's viewpoint, our proposed method naturally extends the A2C framework, and we believe it is also worthwhile to investigate valid designs for future impact decomposition under DDPG or HAC.

\bibliographystyle{ACM-Reference-Format}
\bibliography{main}

\appendix

\section{Analysis on Item Decomposition}\label{ap: example_decomposition}

\subsection{An Intuitive Example}

In this section, we use some examples to show how different item decomposition strategies may influence the learning process.
Considering the relation between request-level and item-level reward:

\begin{equation}
    R(s_t,a_t) = \sum_{i_{t,k} \in a_t} r_{t,k}
\end{equation}

Suppose we observe item-wise click-or-not reward as \\ $[1.0,0.0,0.0,1.0,0.0,0.0]$ for a list of size $K=6$, meaning that only the first and the fourth item got a click, then the total reward of the list is $R(s_t,a_t)=2.0$.
Recall our re-weighting strategy as:
\begin{equation}
    w_{t,k} = \frac{\alpha r_{t,k} + (1-\alpha)}{\alpha R(s_t,a_t) + (1-\alpha)K }
\end{equation}
Considering $\alpha\in \{0, 0.5, 1.0\}$, we have the following re-weighting strategies:
\begin{equation}
\begin{aligned}
    \relax
    [\frac{1}{6},\frac{1}{6},\frac{1}{6},\frac{1}{6},\frac{1}{6},\frac{1}{6}] & \ \ \ \ \text{if } \alpha=0\\
    [\frac{1}{4},\frac{1}{8},\frac{1}{8},\frac{1}{4},\frac{1}{8},\frac{1}{8}] & \ \ \ \ \text{if } \alpha=0.5\\
    [\frac{1}{2}, 0, 0, \frac{1}{2}, 0, 0] & \ \ \ \ \text{if } \alpha=1.0
\end{aligned}
\end{equation}
Note that if using a pure re-weighting strategy with $\alpha=1.0$, the items with clicks will split the future reward $V(s_{t+1})$ in half, and other items with no click signals will not account for any of the future impacts.
And the setting $\alpha=0.5$ will find a smoother solution than this re-weighting strategy and captures the item-wise differences compared to the equal-weight strategy.

\section{Decomposition for critic learning}\label{ap: itema2c-C}

\begin{table*}[ht]\centering
    \caption{Component analysis of ItemA2C where $\alpha = 0.5$ in re-weighting strategy.}
  \centering
  \begin{tabular}{cccccc}
    \toprule
    Dataset  & Performance & ItemA2C & ItemA2C-W(A) & ItemA2C-W(C) & ItemA2C-M(A) \\
     \midrule
   \multirow{2}{*}{ML1M} & Reward  & 17.58$ \pm(0.62)$ & 17.57$ \pm(0.54)$ & 17.57$ \pm(0.54)$ & \textbf{17.94$ \pm(0.47)$} \\
   & Depth  & 17.94$ \pm(0.53)$ & 17.93$ \pm(0.46)$ & 17.92$ \pm(0.46)$ & \textbf{18.24$ \pm(0.40)$} \\

    \midrule
    \multirow{2}{*}{KuaiRand} & Reward  & 13.45$ \pm(0.08)$ & 14.77$ \pm(0.54)$ & 13.41$ \pm(0.40)$ & \textbf{16.03$ \pm(0.53)$}\\
   & Depth  & 14.41$ \pm(0.07)$ & 15.52$ \pm(0.35)$ & 14.38$ \pm(0.45)$ & \textbf{16.59$ \pm(0.45)$} \\
  
    \bottomrule
  \end{tabular}
    
    \label{tab:ab_component}
  
\end{table*}
In our method, the critic is following the TD learning objective of request level optimization where can be easily decomposed by future impact. We provide reward performance of this alternative in the "itema2c-W(C)" column of the following table. It is not even surpassing itemA2C (without reweighting). We believe that this may be related to the fact that we deploy the actor instead of the critic during recommendation. Theoretically, it is possible to decompose the critic learning into item-level, but the items in the same state will eventually aggregate their results under request-level MDP.

\section{The Details of all Methods}\label{ap: model_specification}

All models and baselines use the same user request encoder and the main difference locates in the design of the learning framework and the actor/critic design.
Except for the SlateQ that generate K items simultaneously and optimizes the joint action, all other baselines output a single vector action and adopt KNN-style top-K selection.
We list details of these differences as follows:
\begin{itemize}
\item SlateQ: The framework of SlateQ learning uses TD error on the state-action Q-value function $Q(s_t,i_{t,k})$ and generalizes the "single-choice" assumption:
\begin{equation}
\begin{aligned}
\mathcal{L}_\mathrm{SlateQ} = &\Big(( r_{t,k} + \frac{1}{K} \gamma (1-d) \\ & \sum_{i_{t+1,k^\prime} \in a_{t+1}} Q (s_{t+1},i_{t+1,k^\prime}) - Q(s_t,i_{t,k}) \Big)^2
\label{eq: SlateQ}
\end{aligned}
\end{equation}

\item DDPG: 
The actor learning of Deep DPG method is to maximize $Q(s_t, a_t)$, and the critic learning is derived from the TD error on $Q(s_t, a_t)$, and the actor :
\begin{equation}
\begin{aligned}
\mathcal{L}_\mathrm{critic} &= \left( R(s_t,a_t) + \gamma (1-d)Q(s_{t+1},a_{t+1}) - Q(s_t,a_t) \right)^2\\
\mathcal{L}_\mathrm{actor} &= -Q(s_t,a_t) \text{, where } a_t \sim \pi_\theta (a_t|s_t)
\label{eq: DDPG}
\end{aligned}
\end{equation}

\item Supervision: The supervised learning directly use binary cross entropy loss with real labels. 
\begin{equation}
\begin{aligned}
\mathcal{L}_\mathrm{BCE}   = &- \sum_{i_{t,k} \in a_t} \Big(Y_{i_{t,k}} \log P(i_{t,k}|s_t) \\ & + (1-Y_{i_{t,k}}) \log (1 - P(i_{t,k}|s_t))\Big)
\label{eq: Supervision}
\end{aligned}
\end{equation}
where $Y_{i_{t,k}}$ is the click-or-not label for item $i_{t,k}$.


\item HAC: An embedding vector (i.e. the hyper-action $Z_t$) is used to represent the output list (i.e. the effect-action $a_t$), and the output list is deterministically generated from $Z_t$.
It also includes an inverse module $g$ to infer the hyper-action back from the effect action in order to align the two action spaces through regularization.
The learning framework uses DDPG as the backbone, and includes both the supervision and the action-space alignment to enhance its performance:
\begin{equation}
\begin{aligned}
& \mathcal{L}_\mathrm{critic} = \big( R(s_t,a_t) + \gamma (1-d)Q(s_{t+1},g(Z_{t+1})) - Q(s_t,g(a_t)) \big)^2\\
& \mathcal{L}_\mathrm{actor} =  -A(s_t,Z_t) \log \pi_\theta(a_t|s_t)\\
& \mathcal{L}_\mathrm{hyper} = |Z_t - g(a_t)|^2 \\
& \mathcal{L}_\mathrm{supervision}  = \mathcal{L}_\mathrm{BCE}
\label{eq: HAC}
\end{aligned}
\end{equation}
where $A(s_t,Z_t) = R(s_t,a_t) + \gamma(1-d) Q(s_{t+1},a_{t+1}) - Q(s_t,a_t)$ and the effect-action $a_t$ (i.e. the item list) is deterministically selected from $Z_t$.

\end{itemize}

 \begin{figure*}[ht]
\centering
    \subfigure[Performance in ML1M]{
    \label{ML1M}
  \includegraphics[scale=0.244]{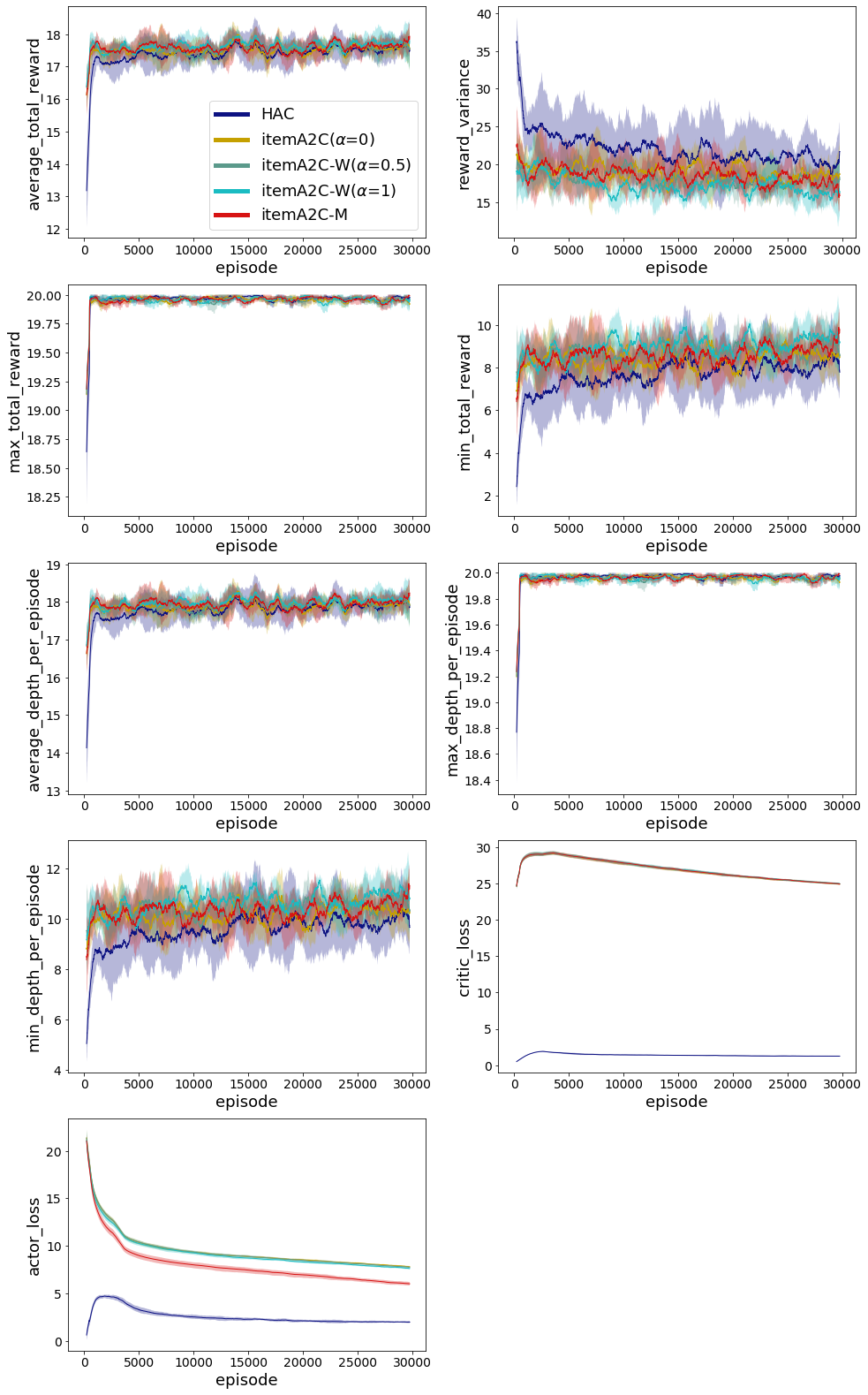}}\subfigure[Performance in KuaiRand]{
    \label{KuaiRand}
  \includegraphics[scale=0.244]{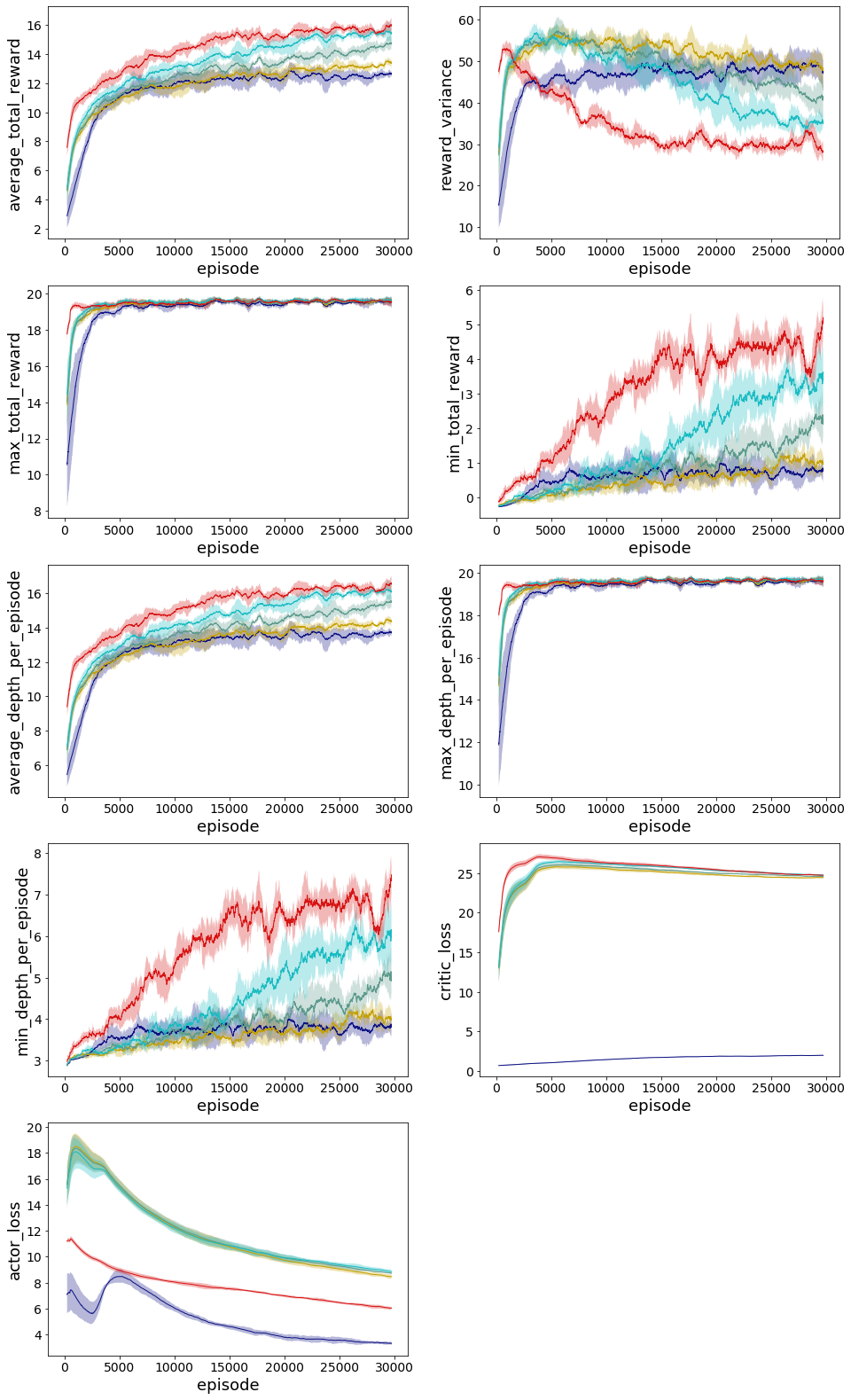}}
\caption{Training curves of all methods. }\label{fig: all_cur}
\centering
\end{figure*}

\section{Full training curves }\label{ap: training_curves}

In Figure \ref{fig: all_cur} we present the training curves of all metrics (reward, variance, and depths) and training losses (critic and actor) in the online learning at both user environments.
We include HAC as the strongest baseline and compare our ItemA2C alternatives with it.
We can see that ItemA2C-M outperforms others on all metrics in the ML1M environment and significantly outperforms others in KuaiRand.
Noted that $\mathcal{L}_\mathrm{actor}$ and $\mathcal{L}_\mathrm{critic}$ of gradually converge during the training process. 
We can also observe the lower reward variance of ItemA2C-M's across users, indicating that the learned policy is more capable of providing good and stable actions under different user states.





\end{document}